\documentclass[twoside]{book}
\usepackage{graphicx}		
\usepackage{url}
\usepackage{psfrag}
\usepackage{amsmath}

 \makeatletter
\ifx\UNDEF\mail\def\mail{ }\else\fi
\ifx\UNDEF\prange\def\prange{0 0}\else\fi

\gdef\@empty{}
\def\Mail#1 #2 {\gdef\thecontact{#1}\gdef\theaddr{#2}}
\def\Range#1 #2 {\gdef\thefirstpage{#1}\gdef\thelastpage{#2}}
{\let\'\mail \expandafter\Mail\' }	
{\let\'\prange \expandafter\Range\' }	
 \gdef\@shtitle{\relax}
 \long\def\shtitle#1{\gdef\@shtitle{#1}}
 \long\def\author#1{\gdef\@author{#1}}
 \def\affil#1{\par\noindent{\rm#1\par}}
 \gdef\@abstract{}
 \long\def\abstract#1{\gdef\@abstract{#1}}

 \renewcommand{\@evenhead}{\thepage\qquad\qquad\@shtitle\hfil}
 \renewcommand{\@oddhead}{\hfil\@shtitle\qquad\qquad\thepage}

 \def\maketitle{\thispagestyle{empty}\chapter{\@title}}
 \renewcommand\chapter{\if@openright\cleardoublepage\else\clearpage\fi
                    \thispagestyle{empty}%
                    \global\@topnum\z@
                    \@afterindentfalse
                    \secdef\@chapter\@schapter}
 \def\@makechapterhead#1{%
  \vspace*{50\p@}%
  {\parindent \z@ \raggedleft \normalfont
    \ifnum \c@secnumdepth >\m@ne
      \if@mainmatter
        \huge 
      \fi
    \fi
    \interlinepenalty\@M
    \Huge \bfseries #1\par\nobreak
    \vskip.25in
    \large\bfseries\@author\par\nobreak
    \vskip 40\p@}
    \ifx\@abstract\@empty\else{\small\@abstract\par\vskip20\p@}\fi
  }

\DeclareRobustCommand\em
        {\@nomath\em \ifdim \fontdimen\@ne\font >\z@
                       \upshape \else \slshape \fi}

\def\@begintheorem#1#2{\sl \trivlist \item[\hskip \labelsep{\bf #1\ #2}]}
\def\@opargbegintheorem#1#2#3{\sl \trivlist
     \item[\hskip \labelsep{\bf #1\ #2\ (#3)}]}

 \newcommand{\sect}[1]{\S\ref{sect:#1}}      
 \newcommand{\eq}[1]{(\ref{eq:#1})}	
 \newcommand{\fig}[1]{Fig.~\ref{fig:#1}}

 \newcommand{\sectlabel}[1]{\label{sect:#1}}
 \newcommand{\eqlabel}[1]{\label{eq:#1}}
 \newcommand{\figlabel}[1]{\label{fig:#1}}

  \def\@arabic#1{\number #1} 

\long\def\@makecaption#1#2{
	\vskip\abovecaptionskip
	\sbox\@tempboxa{{\small {\bf #1}: #2}}%
	\ifdim\wd\@tempboxa>\hsize
	    {\small {\bf #1}: #2\par}
	\else
	   \global\@minipagefalse
	   \hbox to\hsize{\hfil\box\@tempboxa\hfil}
	\fi
	\vskip \belowcaptionskip}

\def\figstrut#1{\hbox to\linewidth{\vrule height#1\hfill}}

\newcommand{\Fig}[4][!htb]{
\begin{figure}[#1]
 \centering\leavevmode#3%
 \caption{#4}
 \figlabel{#2}
\end{figure}                 }

\makeatother

\shtitle{Estimating the dynamics of kernel-based evolving networks}
\title{ Estimating the dynamics of kernel-based evolving networks}
\author{%
  G\'abor Cs\'ardi\affil{Center for Complex Systems Studies, 
    Kalamazoo, MI, USA and Department of Biophysics, KFKI Research
    Institute for Particle and Nuclear Physics of the Hungarian Academy
    of Sciences, Budapest, Hungary\\csardi@kzoo.edu}
  
  Katherine Strandburg\affil{DePaul University -- College of Law, Chicago, IL,
    USA}
  
  L\'aszl\'o Zal\'anyi\affil{Department of Biophysics, KFKI Research
    Institute for Particle and Nuclear Physics of the Hungarian
    Academy of Sciences, Budapest, Hungary}
  
  Jan Tobochnik\affil{Department of Physics and Center for Complex
    Systems Studies, Kalamazoo College, Kalamazoo, MI, USA}
  
  P\'eter \'Erdi\affil{Center for Complex Systems Studies, Kalamazoo College,
    Kalamazoo, MI, USA}
}
\date{May 10, 2006}
\abstract{In this paper we present the application of a novel methodology to
  scientific citation and collaboration networks. This methodology is
  designed for understanding the governing dynamics of evolving
  networks and relies on an \emph{attachment kernel}, a scalar function
  of node properties, which stochastically drives the addition and
  deletion of vertices and edges. We illustrate how the kernel
  function of a given network can be extracted from the history of the
  network and discuss other possible applications.
}

\begin{document}           
\maketitle
                           
\section{Introduction}\sectlabel{intro}	

The network representation of complex systems has been very  successful.
The key to this success is universality in at least two
senses. First, the simplicity of representing complex systems as
networks makes it possible to apply network theory to very different systems,
ranging from the social structure of a group to the interactions of
proteins in a cell. Second, these very different networks show
universal structural traits such as the small-world property and
the scale-free degree-distribution \cite{watts98,barabasi99a}.
See \cite{albert02,newman03} for reviews of complex network research.

Usually it is assumed that the life of most complex systems is defined
by some -- often hidden and unknown -- underlying governing
dynamics. These dynamics are the answers to the question  `How does it work?'
 and a fair share of scientific effort  is taken to uncover this dynamics.

In the network representation the life of a (complex or not)
system is modeled as an evolving graph: sometimes new vertices are
introduced to the system while others are removed, new edges are
formed, others break and all these events are governed by the
underlying dynamics. See \cite{berg04,newman01a,barabasi02a} for data-driven
network evolution studies.

This paper is organized as follows. In Section~\sect{model} we
define a framework for studying the dynamics of two types of evolving
networks and show how this dynamics can be measured from the data.
In Section~\sect{appl} we present two applications and finally
in Section~\sect{discuss} we discuss our results and other possible
applications. 

\section{Modeling evolving networks by 
  attachment kernels}\sectlabel{model}

In this section we introduce a framework in which  the underlying dynamics of evolving networks
can be estimated from knowledge of the time dependence of the evolving network. 

This framework is a discrete time model, where time is measured by the
different events happening in the network. An event is a structural
change: vertex and/or edge additions and/or deletions. The
interpretation of an event depends on the system we're
studying; see Section~\sect{appl} of this paper for two
examples.

The basic assumption of the model is that edge additions depend on some
properties of the vertices of the network. This property can be a
structural one such as the degree of a vertex or its clustering
coefficient but also an intrinsic one such as the age of a person in a
social network or her yearly net income. The model is independent of
the {\em meaning} of these properties.

The vertex properties drive the evolution of the network
stochastically through an attachment kernel, a function giving the
probabilities for any new edges which might be added to the network.
See \cite{krapivsky00} for another possible application of attachment
kernels. 

In this paper we specify the model framework for two special kinds of
networks: citation and non-decaying networks, more general results
will be published in forthcoming publications.

\subsection{Citation networks}

Citation networks are special evolving networks. In a citation network
in each time step (event) a single new node is added to the
network together with its edges (citations). Edges between
``old'' nodes are never introduced and there are no edge or vertex
deletions either.

For simplicity let us assume that the $A(\cdot)$ attachment kernel
depends on a only one property of the potentially cited vertices,
their degree. (The formalism can be generalized easily to
include other properties as well.) We assume that the probability that
at time step $t$ an edge $e$ of a new node will attach to an old node
$i$ with degree $d_i$ is given by 
\begin{equation}
P[\text{$e$ cites $i$}]=\frac{A(d_i(t))}{\sum_{k=1}^{t} A(d_k(t))}
\eqlabel{citdefine}
\end{equation}
The denominator is simply the sum of the attachment kernel functions
evaluated for every node of the network in the current time step. 

With this simple equation the model framework for citation networks is
defined: we assume that in each time step a single new node is
attached to the network and that it cites other, older nodes with the
probability given by \eq{citdefine}.

For a given citation network we can use this model to estimate the
form of the kernel function based on data about the history of a
network. In this paper we only give an overview of this estimation
process, please see \cite{csardi05} for the details.

Based on \eq{citdefine} the probability that an edge $e$ of a new node
at time $t$ cites an old node with degree $d$ is given by
\begin{equation}
P[\text{$e$ cites a $d$-degree node}]=P_e(d)=
  \frac{A(d) N_d(t)}{S(t)}, \qquad S(t)=\sum_{k=1}^{t} A(d_k(t))
\end{equation}
$N_d(t)$ is the number of $d$-degree nodes in the network in time
step $t$. From here we can extract the $A(d)$ kernel function:
\begin{equation}
A(d)=\frac{P_e(d) S(t)}{N_d(t)}
\eqlabel{citest}
\end{equation}
If we know $S(t)$ and $N_d(t)$, then by estimating $P_e(d)$
based on the network data we have an estimate for $A(d)$ via
\eq{citest}, and by doing this for each edge and $d$ degree, in
practice we can have a reasonable approximation of the $A(d)$ function
for most $d$ values. (Of course we cannot estimate $A(d)$ for those
degrees which were never present in the network.)

It is easy to calculate $N_d(t)$ so the only piece missing for the
estimation is that we need $S(t)$ as well; however this is defined in
terms of the measured $A(d)$ function. We
can use an iterative approach to make better and better
approximations for $A(d)$ and $S(t)$. First we assume that $S_0(t)=1$
for each~$t$ and measure $A_0(d)$ which can be used to calculate the
next approximation of $S(t)$, $S_1(t)$ yielding $A_1(t)$ via the
measurement, etc. In practice this procedure converges quickly for the
systems we have studied -- after
five iterations the difference between successive $A_n(d)$ and
$A_{n+1}(d)$ estimations is very small.

\subsection{Non-decaying networks}

Non-decaying networks are more general then citation networks because
connections can be formed between older nodes as well. It is still
true, however, that neither edges nor nodes are ever removed from the
network.

Similarly to the previous section, we assume that the attachment
kernel depends on the degree of the vertices, but this time on the
degree of both vertices involved in the potential connection. The
probability of forming an edge between nodes $i$ and $j$ in time step
$t$ is given by 
\begin{equation}
P[\text{$i$ and $j$ will be connected}]=\frac{A(d_i(t),d_j(t))}
  {\sum_{k}^{N(t)}\sum_{l\ne k}^{N(t)} (1-a_{kl}(t))A(d_k(t), d_l(t))}
\end{equation}
The denominator is the sum of the attachment kernel function applied to all
possible (not yet realized) edges in the network. $a_{kl}(t)$ is 1
if there is an edge between nodes $k$ and $l$ in time step $t$ and
0 otherwise.

Using an argument aimilar to that of the previous section we can estimate
$A(d^*, d^{**})$ via
\begin{equation}
P[\text{$e$ connects $d^*$ and $d^{**}$ degree nodes}]=P_e(d^*,d^{**})=
  \frac{A(d^*, d^{**}) N_{d^*,d^{**}}(t)}{S(t)},
\end{equation}
$N_{d^*,d^{**}}(t)$ is the number of not yet realized edges between $d^*$
and $d^{**}$ degree nodes in time step $t$, and 
\begin{equation}
S(t)=\sum_{k}^{N(t)}\sum_{l\ne k}^{N(t)} (1-a_{kl}(t))A(d_k(t), d_l(t))
\end{equation}
\begin{equation}
A(d^*, d^{**})=\frac{P_e(d^*, d^{**}) S(t)}{N_{d^*,d^{**}}(t)}
\end{equation}

S(t) can be approximated using an iterative approach similar to that
introduced in the previous section. 

\section{Applications}\sectlabel{appl}

In this section we briefly present results for two applications for the model
framework and measurement method. For other applications and details
see \cite{csardi05,csardi06}.

\subsection{Preferential attachment in citation networks}

The preferential attachment model \cite{barabasi99a,jeong03} gives a mechanism
to generate the scale-free degree-distribution often found in various
networks. In our framework for citation networks it simply means that
the kernel function linearly depends on the degree: 
\begin{equation}
  A(d)=d+a,
\end{equation}
where $a$ is a constant.

By using our measurement method, it is possible to measure the kernel
function based on node degree for various citation networks and check
whether they evolve based on this simple principle.

Let us first consider the network of high-energy physics papers from
the arXiv e-print archive. We used data for papers submitted between
January, 1992 and July, 2003, which included 28632 papers and 367790
citations among them. The data is available online at
\url{http://www.cs.cornell.edu/projects/kddcup/datasets.html}.
This dataset and other scientific citation networks are well studied,
see \cite{redner05,lehmann03} for examples.

First we've applied the measurement method based on the node degree 
to this network and found that indeed, the attachment kernel of the
network is close to the one predicted by the preferential attachment
model, that is 
\begin{equation}
A^*_\text{HEP}(d)=d^{0.85}+1
\end{equation}
gives a reasonably good fit to the data. See the measured form of
the kernel in \fig{hepdeg}.

The small exponent for $d$ is in good agreement
with the fact that the degree distribution of this network decays
faster than a power-law. 

Next, we've applied the measurement method by using two properties of
the potentially cited nodes: their degree and age, the latter is
simply defined as the difference of the current time step and the time step
when the node was added. We found that the two variable $A(d,a)$
attachment kernel has the following form:
\begin{equation}
A^{**}_\text{HEP}(d,a)=(d^{1.14}+1) \, a^{-1.14}.
\end{equation}
This two-variable attachment kernel gives a better understanding of
the dynamics of this network: the citation probability increases about
linearly with the degree of the nodes and decreases as a power-law
with their age. Note that these two effects were both present in the
degree-only dependent $A^*$ attachment kernel, this is why the
preferential attachment exponent was smaller there ($0.85<1.14$).

\Fig{hepdeg}{%
  \begin{psfrags}
  \psfrag{mark-0000}[Bc][Bc]{0}
  \psfrag{mark-0001}[Bc][Bc]{50}
  \psfrag{mark-0002}[Bc][Bc]{100}
  \psfrag{mark-0003}[Bc][Bc]{150}
  \psfrag{mark-0004}[Bc][Bc]{200}
  \psfrag{mark-0005}[Bc][Bc]{degree}
  \psfrag{mark-0006}[Bc][Bc]{$A^*_\text{HEP}(d)$}

  \includegraphics[width=0.48\textwidth]{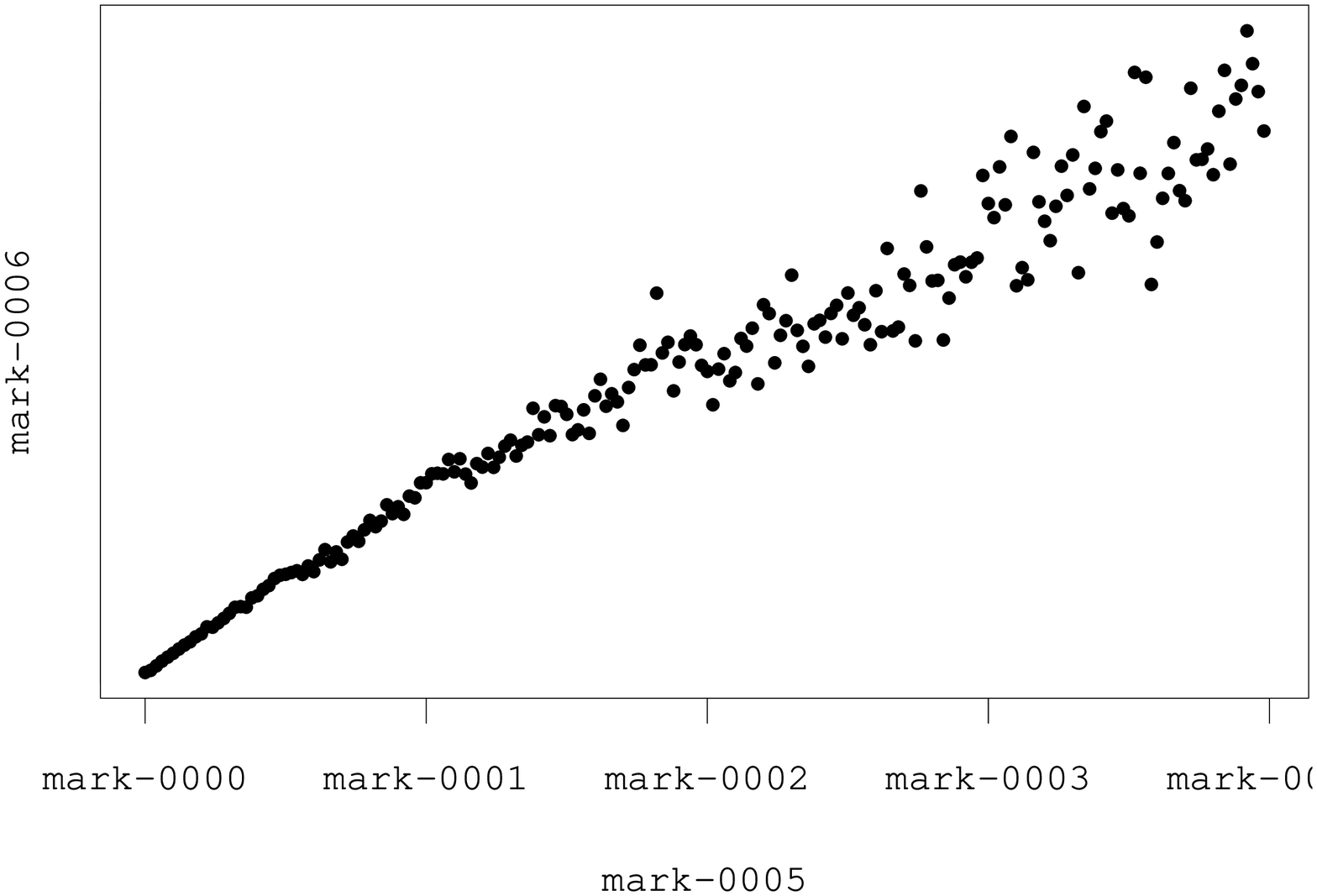}
  \end{psfrags}
\hfill%
  \begin{psfrags}
  \psfrag{mark-0000}[Bc][Bc]{1}
  \psfrag{mark-0001}[Bc][Bc]{2}
  \psfrag{mark-0002}[Bc][Bc]{5}
  \psfrag{mark-0003}[Bc][Bc]{10}
  \psfrag{mark-0004}[Bc][Bc]{20}
  \psfrag{mark-0005}[Bc][Bc]{degree}
  \psfrag{mark-0006}[Bc][Bc]{$A^{**}_\text{HEP}(d^*,d^{**})$}
  \psfrag{mark-0007}[Bl][Bl]{10}
  \psfrag{mark-0008}[Bl][Bl]{20}
  \psfrag{mark-0009}[Bl][Bl]{30}
  \psfrag{mark-0010}[Bl][Bl]{40}

  \includegraphics[width=0.48\textwidth]{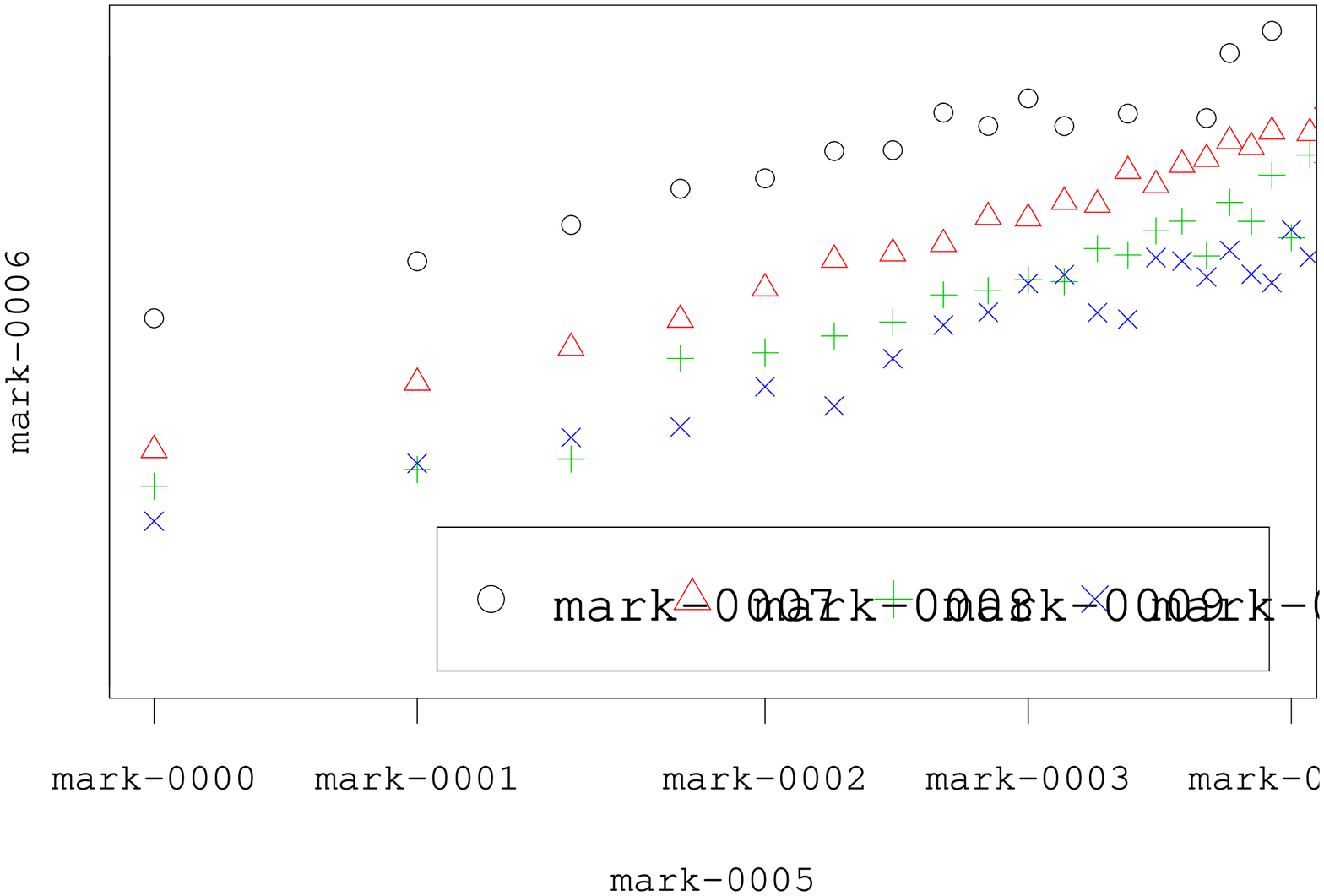}
  \end{psfrags}

}{
  The two measured kernel functions for the HEP citation network. The
  left plot shows the degree dependent kernel, the right the degree and age
  dependent kernel. On the right plot four sections along the degree
  axis are shown for different vertex ages.
}

Similar results were obtained for the citation network of US
patents granted between 1975 and 1999 containing 2,151,314
vertices and 10,565,431 edges: 
\begin{equation}
A^*_\text{patent}(d)=d^{0.79}, \qquad 
A^{**}_\text{patent}(d,a)=(d^{1.2}+1) \, a^{-1.6}.
\end{equation}

These two studies show that the preferential attachment phenomenon can
be present in a network even if it does not have power-law
degree-distribution because there is another process -- {\em aging} in
our case -- which prevents nodes from gaining very many edges. 

\subsection{The dynamics of scientific collaboration networks}

In this section we briefly present the results of applying our methods
to a non-decaying network: the cond-mat collaboration network. 
In this network a node is a researcher who published at least one
paper in the arXiv cond-mat archive between 1970 and 1997 (this is the
date when the paper was submitted to cond-mat, not the actual
publication date, but most of the time these two are almost the
same). There is an edge between two researchers/nodes if they've
published at least one paper together. The data set contains 23708
papers, 17636 authors and 59894 edges. 

We measured the attachment kernel for this network based on the
degrees of the two potential neighbors. See \fig{cond1}
for the $A_\text{cond-mat}(d^*, d^{**})$ function.

\Fig{cond1}{%
  \begin{psfrags}
  \psfrag{mark-0000}[Bc][Bc]{5}
  \psfrag{mark-0001}[Bc][Bc]{10}
  \psfrag{mark-0002}[Bc][Bc]{15}
  \psfrag{mark-0003}[Bc][Bc]{20}
  \psfrag{mark-0004}[Bc][Bc]{25}
  \psfrag{mark-0005}[Bc][Bc]{30}
  \psfrag{mark-0006}[Bc][Bc]{30}
  \psfrag{mark-0007}[Bc][Bc]{25}
  \psfrag{mark-0008}[Bc][Bc]{20}
  \psfrag{mark-0009}[Bc][Bc]{15}
  \psfrag{mark-0010}[Bc][Bc]{10}
  \psfrag{mark-0011}[Bc][Bc]{10}
  \psfrag{mark-0012}[Bc][Bc]{40}
  \psfrag{mark-0013}[Bc][Bc]{30}
  \psfrag{mark-0014}[Bc][Bc]{20}
  \psfrag{mark-0015}[Bc][Bc]{$A_\text{cond-mat}(d^*, d^{**})$}
  \psfrag{mark-0016}[Bc][Bc]{degree}
  \psfrag{mark-0017}[Bc][Bc]{degree}
  \psfrag{mark-0018}[Bc][Bc]{5}

  \includegraphics[width=0.45\textwidth]{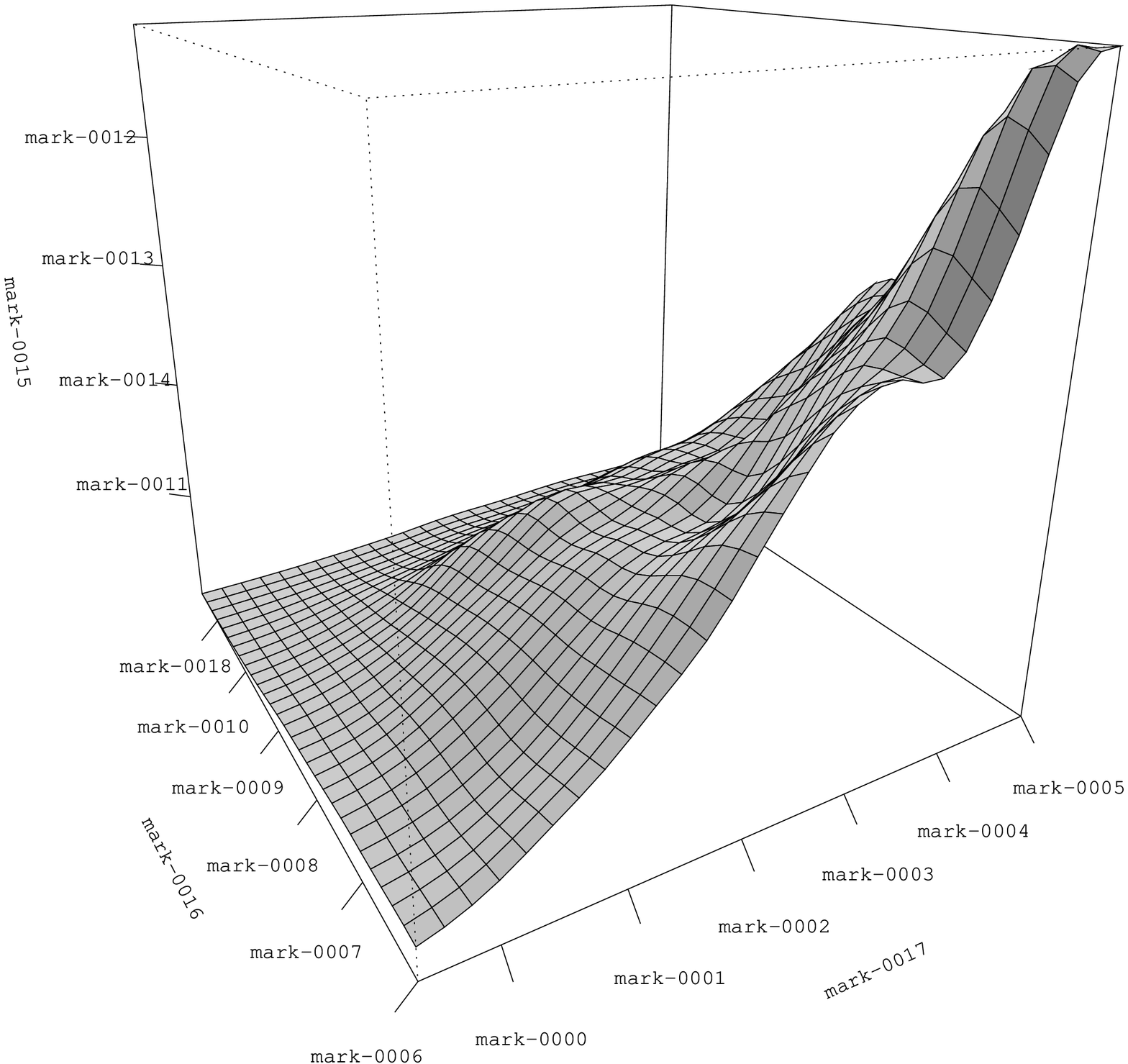}
  \end{psfrags}
\hfill%
  \begin{psfrags}
  \psfrag{mark-0000}[Bc][Bc]{1}
  \psfrag{mark-0001}[Bc][Bc]{2}
  \psfrag{mark-0002}[Bc][Bc]{5}
  \psfrag{mark-0003}[Bc][Bc]{10}
  \psfrag{mark-0004}[Bc][Bc]{20}
  \psfrag{mark-0005}[Bc][Bc]{50}
  \psfrag{mark-0006}[Bc][Bc]{100}
  \psfrag{mark-0007}[Bc][Bc]{degree}
  \psfrag{mark-0008}[Bc][Bc]{0.05}
  \psfrag{mark-0009}[Bc][Bc]{0.5}
  \psfrag{mark-0010}[Bc][Bc]{0.0005}
  \psfrag{mark-0011}[Bl][Bl]{0}
  \psfrag{mark-0012}[Bl][Bl]{1}
  \psfrag{mark-0013}[Bl][Bl]{10}
  \psfrag{mark-0014}[Bl][Bl]{20}

  \includegraphics[width=0.49\textwidth]{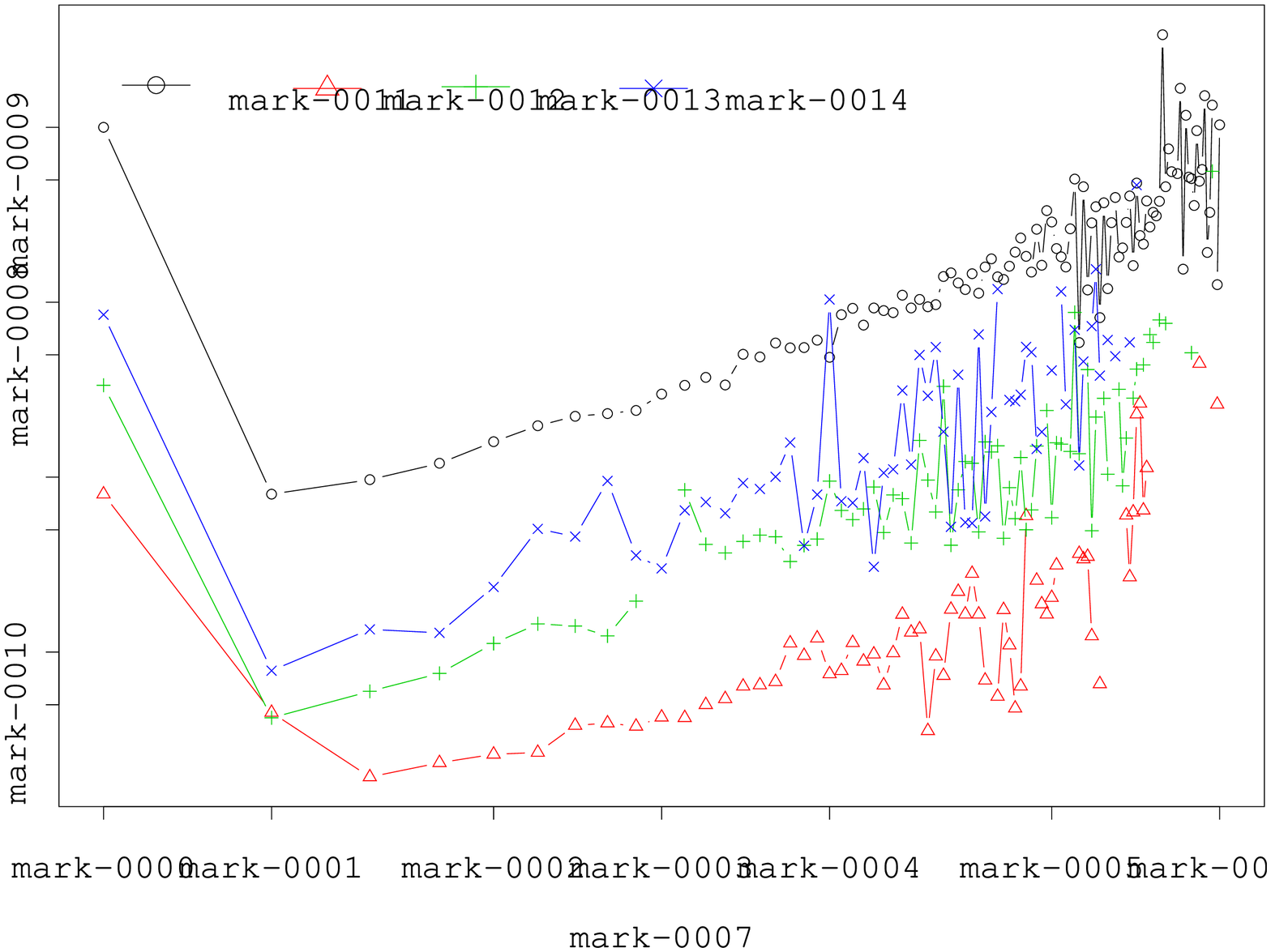}
  \end{psfrags}

}{
  The attachment kernel for the cond-mat collaboration network, the
  surface plot was smoothed by applying a double exponential smoothing
  kernel to it. The right plot has logarithmic axes. The right plot
  shows that the kernel function has high values for zero-degree
  nodes, this might be because a new researcher will usually write a paper 
  with collaborators and thus will have a high probability of adding
  links to the network.
}

We've tried to fit various functional forms to the two-dimensional
attachment kernel function to check which is a better description of
the dynamics. See \fig{condfits} for the shape of the fitted
functions and Table~\ref{tab:constable} for the functional forms and
the results. 

The best fit was obtained by 
\begin{equation}
A'_\text{cond-mat}(d^*,d^{**})=c_1\cdot (d^* d^{**})^{c_2} + c_3
\end{equation}
where the  $c_i$ are constants.

See \cite{barabasi02a,newman01} for other studies on collaboration
networks.

\Fig{condfits}{%
  \begin{psfrags}
  \psfrag{mark-0000}[Bc][Bc]{{\large\bf A}}

  \includegraphics[width=0.4\textwidth]{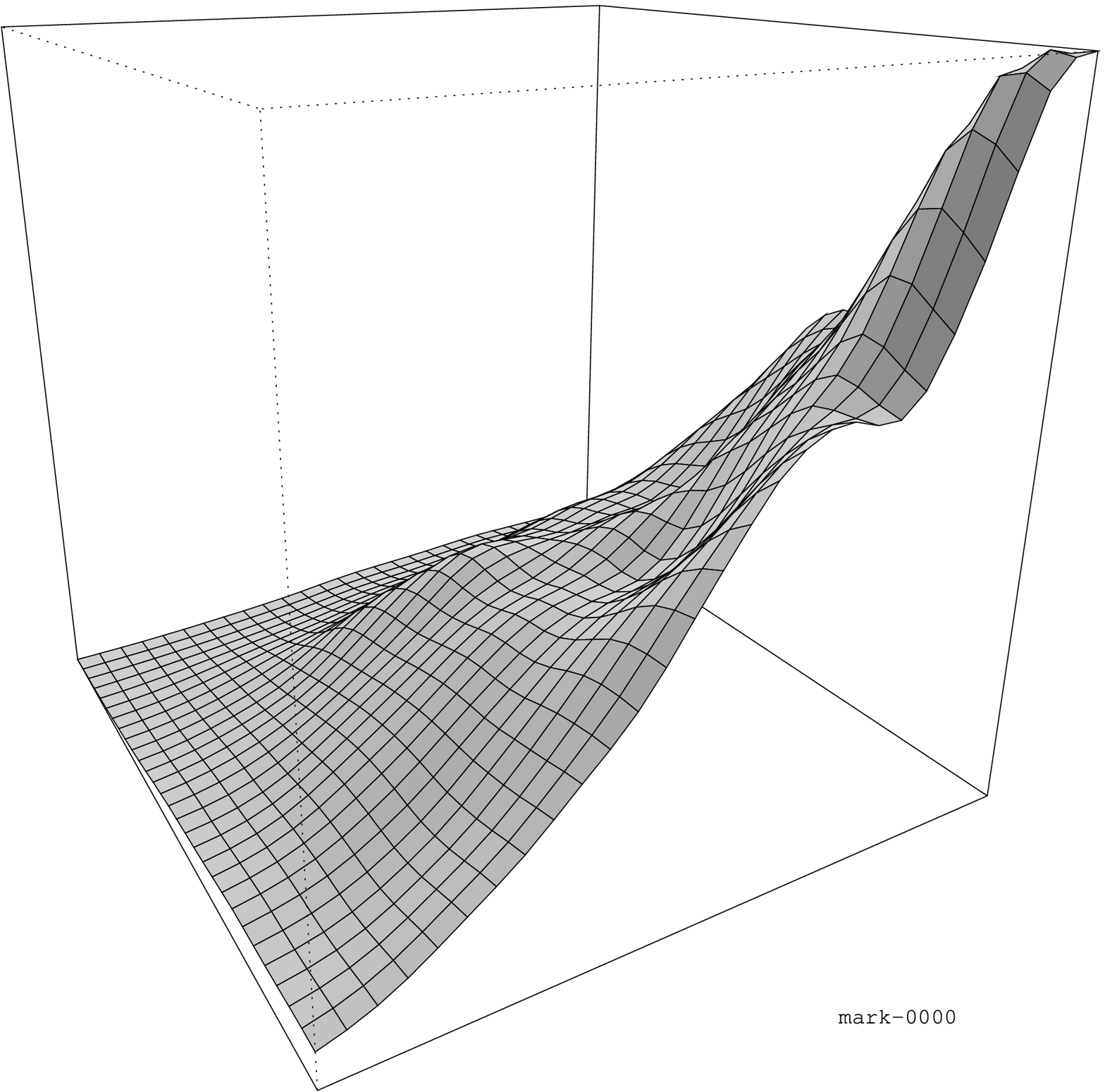}
  \end{psfrags}
\hfill%
  \begin{psfrags}
  \psfrag{mark-0000}[Bc][Bc]{{\large\bf B }}

  \includegraphics[width=0.4\textwidth]{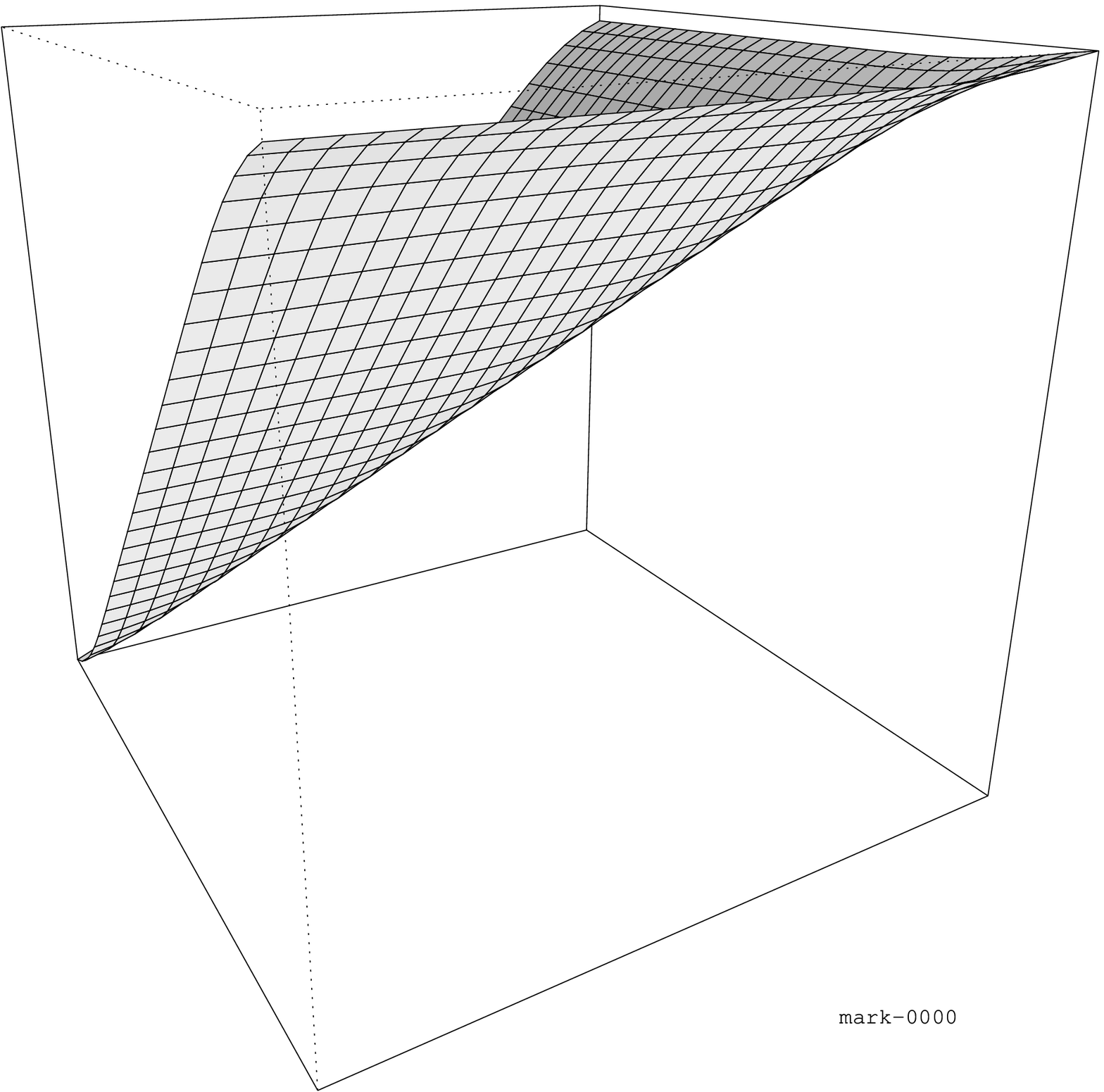}
  \end{psfrags}
\\
  \begin{psfrags}
  \psfrag{mark-0000}[Bc][Bc]{{\large\bf C }}

  \includegraphics[width=0.4\textwidth]{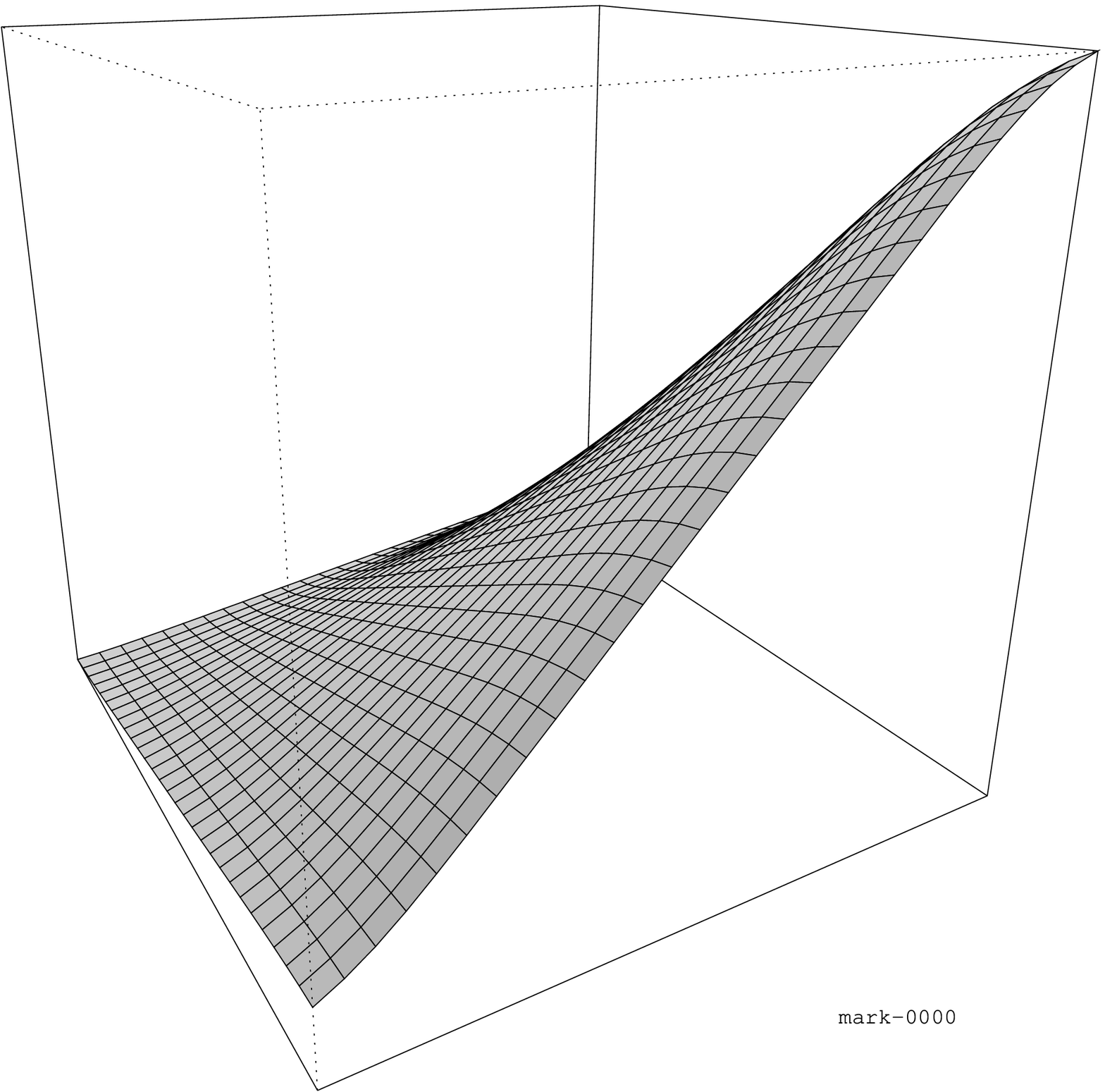}
  \end{psfrags}
\hfill%
  \begin{psfrags}
  \psfrag{mark-0000}[Bc][Bc]{{\large\bf D }}

  \includegraphics[width=0.4\textwidth]{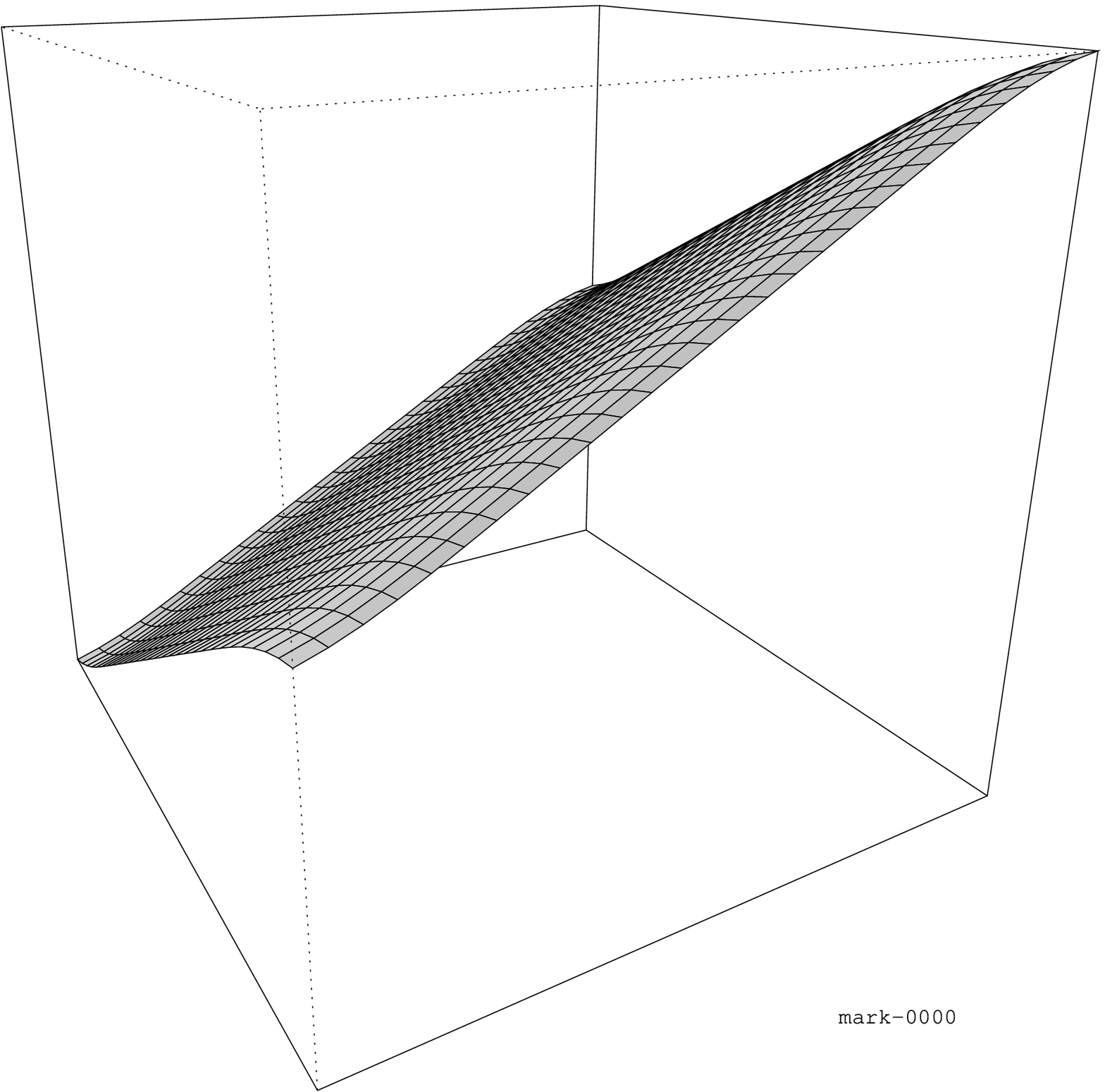}
  \end{psfrags}
\\
}{
  \textbf{A} shows the smoothed measured kernel function for the
  collaboration network, \textbf{B}, \textbf{C} and \textbf{D} are
  fitted functional forms shown in the first three lines of
  Table~\ref{tab:constable}. The best fit is clearly obtained by the
  multiplicative fit.
}

\begin{table}
  \renewcommand{\arraystretch}{1.5}
  \begin{tabular}{ll|l|r|l}
    & Fitted form & Fitted parameters & Fit Error & Fitting method \\
    \hline
    \textbf{B} & 
    $c_1\max(d^*,d^{**})+c_2$ & $c_1=1.26$, $c_2=-10.56$ & 107357.6 &
        Nelder-Mead \\
    \textbf{C} &
    $c_1d^*d^{**}+c_2$ & $c_1=0.0697$, $c_2=-2.11$ & 4300.2 & 
        Nelder-Mead \\
    \textbf{D} & 
    $c_1(d^*+d^{**})+c_2$ & $c_1=1.08$, $c_2=-18.98$ & 31348.9 & 
        Nelder-Mead \\\hline
    & $c_1d^*d^{**}+c_2(d^*+d^{**})+$ & $c_1=0.0783$, $c_2=-0.12$,
        & 3532.9 & BFGS \\[-3pt]
    & \ \ \ \ $+c_3\max(d^*,d^{**})+c_4$ & $c_3=-0.093$, $c_4=1.50$ & & \\
    & $c_1(d^*d^{**})^{c_2}+c_3$ & $c_1=0.016$, $c_2=1.22$, &
        3210.4 & SANN \\
    & & $c_3=0.58$ & & \\
  \end{tabular}
  \caption{Four optimization methods were run for each functional
    form to minimize the least square difference: BFGS, Nelder-Mead,
    CG and SANN, the results of the best fits are included in the
    table. See \cite{nocedal99,belisle92} for the details of these
    methods.}
  \label{tab:constable}
\end{table}

\section{Discussion}\sectlabel{discuss}

We've briefly presented a methodology for understanding the evolution of
networks through kernel functions and showed how the kernel functions
can be extracted from network data. 

We've discussed two applications for this methodology: first the
``fitting'' of the preferential attachment model to a network of
scientific citations and then determining how the evolution of a
scientific collaboration network depends on the degree of the
vertices. 

The methodology outlined here is general and can be
successfully applied to \emph{any} kind of evolving network where time dependent data is available.
By defining the kernel function in
terms of the potentially important vertex properties one can check
whether these properties really significantly influence network
evolution: if a kernel function is not sensitive to one of its
arguments that suggests that this argument does not have an important
contribution.
Another possible application would be to identify changes
in the dynamics of a system by doing the measurements in sliding time
windows, see \cite{csardi06} for an example.

\section{Acknowledgement}

This work was funded in part by the EU FP6 Programme
under grant numbers IST-4-027173-STP and IST-4-027819-IP and by the
Henry R. Luce Foundation. The authors also thank Mark Newman for
providing the cond-mat data set.

\clearpage
\bibliography{../../patentnet}

\end{document}